# The Easiest Hard Problem: Number Partitioning

Stephan Mertens*
*Inst. f. Theor. Physik, Univ. Magdeburg, Universitätsplatz 2, 39108 Magdeburg, Germany*

Number partitioning is one of the classical NP-hard problems of combinatorial optimization. It has applications in areas like public key encryption and task scheduling. The random version of number partitioning has an "easy-hard" phase transition similar to the phase transitions observed in other combinatorial problems like $k$-SAT. In contrast to most other problems, number partitioning is simple enough to obtain detailled and rigorous results on the "hard" and "easy" phase and the transition that separates them. We review the known results on random integer partitioning, give a very simple derivation of the phase transition and discuss the algorithmic implications of both phases.

## I. INTRODUCTION

The number partitioning problem (NPP) is defined easily: Given a list $a_1, a_2, \ldots, a_N$ of positive integers, find a partition, i.e. a subset $\mathcal{A} \subset \{1, \ldots, N\}$ such that the discrepancy

$$E(\mathcal{A}) = \Big| \sum_{i \in \mathcal{A}} a_i - \sum_{i \notin \mathcal{A}} a_i \Big|, \tag{1}$$

is minimized. A partition with $E = 0$ ($E = 1$) for $\sum a_j$ even (odd) is called *perfect partition* for obvious reasons.

Number partitioning is of considerable importance, both practically and theoretically. Its practical applications range from multiprocessor scheduling and the minimization of VLSI circuit size and delay [1, 2] over public key cryptography [3] to choosing up sides in a ball game [4]. Number partitioning is also one of Garey and Johnson's six basic NP-hard problems that lie at the heart of the theory of NP-completeness [5, 6], and in fact it is the only one of these problems that actually deals with numbers. Hence it is often chosen as a base for NP-completeness proofs of other problems involving numbers, like bin packing, multiprocessor scheduling [7], quadratic programming or knapsack problems.

The computational complexity of the NPP depends on the type of input numbers $\{a_1, a_2, \ldots, a_N\}$. Consider the case that the $a_j$'s are positive integers bound by a constant $A$. Then the discrepancy $E$ can take on at most $NA$ different values, i.e. the size of the search space is $\mathcal{O}(NA)$ instead of $\mathcal{O}(2^N)$ and it is straightforward to devise an algorithm that explores this reduced search space in time polynomial in $NA$ [5]. Unfortunately, such an algorithm does not prove P=NP since a concise encoding of an instance requires $\mathcal{O}(N \log A)$ bits, and $A$ is not bounded by any polynomial of $\log A$. This feature of the NPP is called "pseudo polynomiality". The NP-hardness of the NPP requires input numbers of size exponentially large in $N$ or, after division by the maximal input number, of exponentially high precision.

To study typical properties of the NPP, the input numbers are usually taken to be independently and identically distributed random numbers. Under this probabilistic assumption, the minimal discrepancy $E_1$ is a stochastic variable. For real valued input numbers (infinite precision, see above), Karmarkar, Karp, Lueker and Odlyzko [8] proved that the *median* value of $E_1$ is $\mathcal{O}(\sqrt{N} \cdot 2^{-N})$ and Lueker [9] showed that the same scaling holds for the *average* value of $E_1$. From numerical simulations [10] it is known that the variance of $E_1$ is of the same order of magnitude as the average, i.e. $E_1$ is *non self averaging.*

Another surprising feature of the NPP is the *poor quality* of heuristic algorithms [11, 12]. The differencing method (see below) is the best polynomial time heuristics known to date, and for real valued $a_j$ it yields discrepancies $\mathcal{O}(N^{-\alpha \log N})$ for some positive constant $\alpha$ [13]. This is far above the true optimum, yet it is the best one can get for large systems! The poor quality of polynomial time heuristics is a very peculiar feature that distinguishes the NPP from many other hard optimization problems like the TRAVELLING SALESMAN PROBLEM [14] for which satisfying approximative algorithms do exist.

The NP-hardness of the NPP tells us that for numbers $a_j$ bounded by $A = 2^{\kappa N}$ the worst case complexity of any exact algorithm is exponential in $N$ for all $\kappa > 0$. Numerical simulations show that the *typical complexity* on instances of the random ensemble depends on $\kappa$. It is exponential and essentially independent of $\kappa$ for $\kappa > \kappa_c > 0$. For $\kappa < \kappa_c$ it is still exponential, but with a base that decreases with decreasing $\kappa$. The critical value $\kappa_c$ marks a transition point where the random ensemble somehow changes its character. Below $\kappa_c$ typical instances seem to have a special property that can be exploited by an exhaustive algorithm. It turns out that this property is the probability of having a perfect partition, which jumps from 0 to 1 as $\kappa$ crosses $\kappa_c$ from above. This abrupt change of quantity is called a *phase transition* in analogy to the transitions observed in thermodynamic systems. Phase transitions in average complexity have been observed in many NP-hard problems like SATISFIABILITY [15, 16] or HAMILTONIAN CIRCUIT [17]. Their study forms the base of an emerging interdisciplinary field of research that joints computer scientists, mathematicians and physicists [18, 19]. The transition in the NPP illu-minates the interdisciplinary character of the field. Fu [20] (physicist) mapped partitioning to an infinite-range, antiferromagnetic spin glass and concluded (incorrectly) that this model did not have a phase transition. Gent and Walsh [21] (computer scientists) verified the phase transition by numerical simulations. They introduced the control parameter $\kappa$ and estimated the transition point close to $\kappa_c = 0.96$. Mertens [22] (physicist) reconsidered Fu's spin glass analogy and derived a phase transition at $\kappa_c = 1 - \frac{\log_2 N}{2N} + \mathcal{O}(\frac{1}{N})$. Then Borgs, Chayes and Pittel [23] (mathematicians) took over and established the phase transition and its characterization rigorously. The mathematical proofs for the phase transitions are another exceptional feature of the NPP. For other NP-hard problems like SATISFIABILITY much less is known rigorously and the sharpest results have been obtained by the powerfull, but non-rigorous techniques of statistical mechanics [24, 25].

*E-mail:stephan.mertens@physik.uni-magdeburg.de

It is this combination of algorithmic hardness and analytic tractability that characterizes the NPP as the "easiest hard problem", a phrase coined by Brian Hayes [4]. In this contribution we will exploit the easiness of the NPP to provide an understanding of some of its remarkable properties.

## II. ALGORITHMS AND COMPLEXITY

In view of the NP-hardness of the NPP it is wise to abandon the idea of an exact solution and to ask for an approximative but fast heuristic algorithm. An obvious approach is place the largest number in one of the two subsets. Then continue to place the largest number of the remaining numbers in the subset with the smaller total sum thus far until all numbers are assigned. The idea behind this *greedy heuristics* is to keep the discrepancy small with every decision. The worst that could happen is that the two subsets are perfectly balanced just before the last number has to be assigned. This is the motivation for assigning the numbers in decreasing order, and it gives the scaling of the result: $\mathcal{O}(N^{-1})$ for real-valued $a_j$. This is extremely bad compared to $\mathcal{O}(\sqrt{N}\,2^{-N})$ of the optimum discrepancy. The time complexity of the greedy algorithm is given by the time complexity to sort $N$ numbers, i.e. it is $\mathcal{O}(N \log N)$. Applied to the set $\{a_j\} = \{8,7,6,5,4\}$ the greedy heuristics misses the perfect solution and yields a partition $\{8,5,4\}\,\{7,6\}$ with discrepancy 4.

The differencing method of Karmarker and Karp [26], also called the KK heuristics, is another polynomial time approximation algorithm. The key idea of this algorithm is to reduce the size of the numbers. This is achieved by replacing the two largest numbers by the absolute value of their difference. This differencing operation is equivalent to commit placing both numbers in different subsets without actually fixing the subset each will go in. With each differencing operation the number of numbers decreases by one, and the last number is the final discrepancy. Applied to $\{8,7,6,5,4\}$, the differencing method yields a discrepancy of 2 that results from the partition $\{8,6\}\,\{7,5,4\}$. Note that the reconstruction of the partition requires some extra bookkeeping that we did not mention on our brief description of the differencing method. Again the heuristic algorithm misses the perfect solution, but the outcome is at least better than the greedy result. Yakir [13] proved that the result of the differencing method on random real valued $a_j \in [0,1]$ is $\mathcal{O}(N^{-\alpha \log N})$ with a constant $\alpha = 0.72$. Again this is much better than the greedy result, yet it is far away from the optimum. The time complexity of the differencing method is dominated by the complexity of selecting the two largest numbers. This is most efficiently done by sorting the initial list and keeping the order througout all iterations, leaving us with a time complexity $\mathcal{O}(N \log N)$.

Both heuristics can be used as a base for an exact algorithm. At each iteration, the greedy algorithm decides to place a number in the subset with the smaller total sum so far. The only alternative is to place the number in the other subset. Exploring both alternatives means searching a binary tree that contains all $2^N$ possible partitions. The corresponding alternative in the KK heuristics is to replace the two largest numbers by their sum. Korf [27] calls the algorithms that explore both alternatives *complete greedy* and *complete differencing* algorithm. Fig. 1 shows the search tree of the complete differencing method for our example $\{8,7,6,5,4\}$.

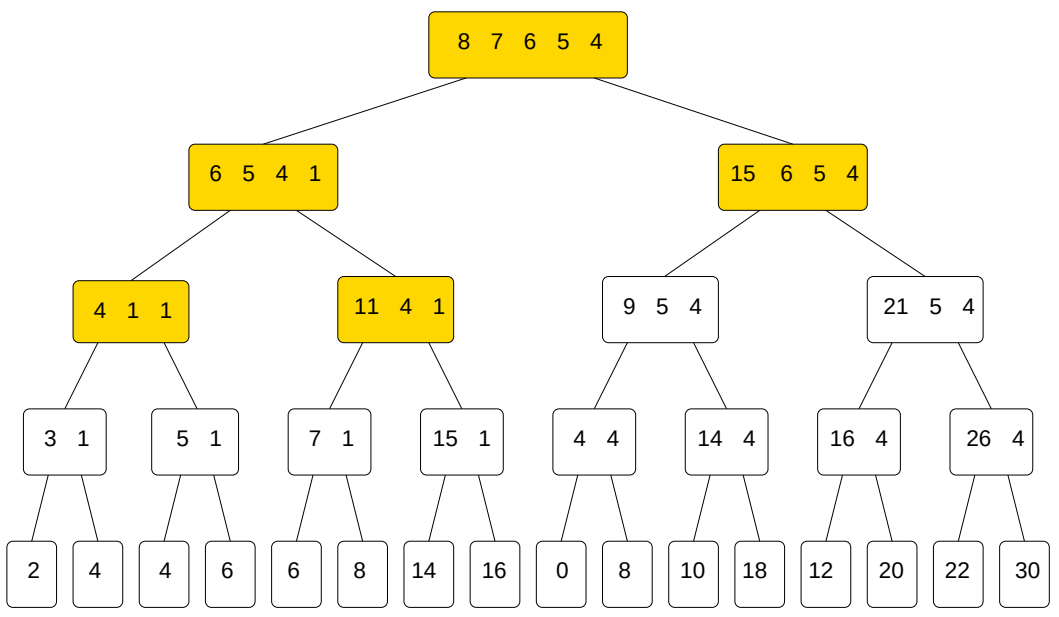

FIG. 1: Search tree of the complete differencing algorithm. Left branch means "replace two largest numbers by their difference", right branch means "replace them by their sum". With appropriate pruning rules only the colored nodes have to be visited to find the optimum solution.

Both complete algorithms have exponential time complexity in the worst case, but it is possible to prune parts of the search tree by simple rules. For the complete differencing method these rules are:

1. If less than 5 numbers are left, take the left branch (i.e. apply the differencing operation).

2. If the largest number in the set set is larger than or equal to the sum of all the other numbers, stop branching: the best solution in this subtree is to place the largest number in one set, all the other numbers in the other set.

3. If a perfect partition has been found, stop the whole algorithm.

The first rule needs some thought, but it can in fact be proven that the KK-heuristics always yields the optimum for $N \leq 4$. Similar pruning rules can be added to the complete greedy method. Fig. 1 shows that the rules really chop off large parts of the search tree, at least in our example.

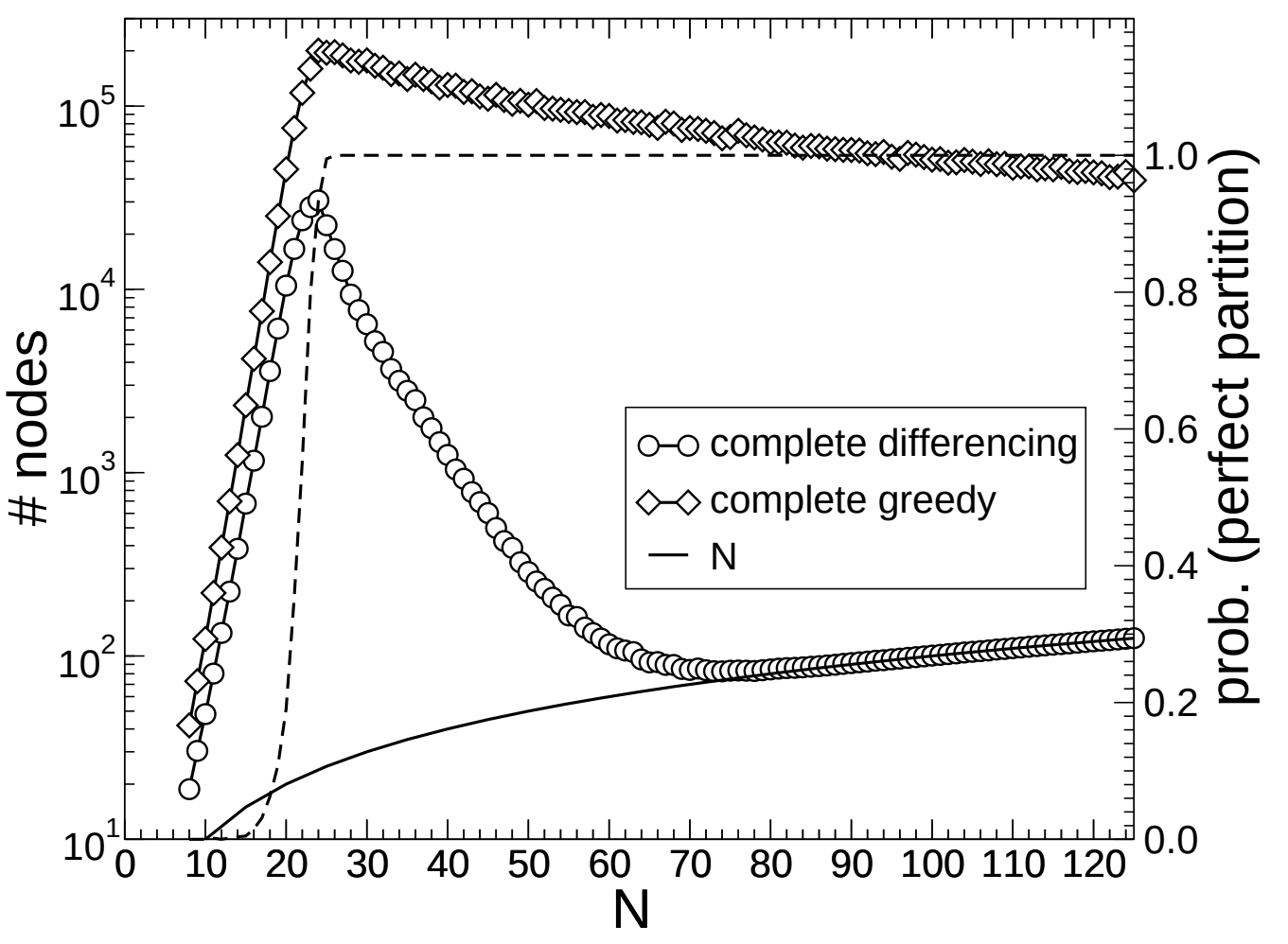

FIG. 2: Number of nodes visited by the complete greedy and the complete differencing algorithm. Instances are sets of random 20-bit integers of cardinality $N$, each data point represents an average over $10^4$ instances. The dashed line indicates the empirical probability that a given instance has a perfect solution.

The question is how the pruning affects the search in general and for large instances. Fig. 2 shows the number of nodes visited by the complete greedy and the complete differencing algorithm for solving large instances of random 20-bit integers. For small values of $N$, the number of nodes growth exponentially with $N$, i.e. the pruning shows only little effect on the performance. For systems beyond $N = 23$ the situation changes drastically: the number of nodes not only stops increasing with $N$, it *decreases*. Larger problems become easier to solve! Apparently the pruning gets more and more effective as $N$ increases, especially for the complete differencing algorithm. For $N > 80$ the latter explores only $N$ nodes of the search tree, i.e. the very first leaf of the tree represents the optimum solution, and the algorithm "knows" that without exploring the rest! This can only mean that rule 3 from above applies, i.e. the partition generated by the differencing heuristic must be perfect.

The appearance of perfect partitions is closely related to the transition in the average complexity, as can be seen from the probability that a random instance has a perfect partition: This probability jumps precisely at the point where the algorithmic complexity changes its behavior, see Fig. 2. Apparently there is computational hard regime without perfect partitions and a computational easy regime where perfect partitions are abundant.

## III. PHASE TRANSITION

As we have seen in the preceeding section the average complexity of algorithms for the random NPP depends on the presence of perfect partitions. The probability of perfect solutions is a property of the ensemble of instances and it can be studied independently from algorithms. This is what we do in this section.

A partition $\mathcal{A}$ can be encoded by binary variables $s_j = \pm 1$: $s_j = +1$ if $j \in \mathcal{A}$, $s_j = -1$ otherwise. The cost function then reads $E = |D(s)|$ where

$$D = \sum_{i=1}^{N} a_i s_i \tag{2}$$

is the signed discrepancy. An alternative cost function is $H = D^2$ or

$$H = -\sum_{i,j} J_{ij} s_i s_j \qquad \text{with } J_{ij} = -a_i a_j. \tag{3}$$

$H$ is the Hamiltonian of an infinite range, antiferromagnetic spin glas, which has been studied by physicists at least three times [10, 20, 22] within the canonical framework of statistical mechanics. Here we follow another, very simple approach that has been used recently to analyze the MULTIPROCESSOR SCHEDULING PROBLEM [7].

The signed discrepancy $D$ can be interpreted as the distance to the origin of a walker in one dimension who takes steps to the left ($s_j = -1$) or to the right ($s_j = +1$) with random stepsizes ($a_j$). The average number of walks that end at $D$ reads

$$\Omega(D) = \sum_{\{s_j\}} \left\langle \delta\left(d - \sum_{j=1}^{N} a_j s_j\right)\right\rangle \tag{4}$$

where $\langle\cdot\rangle$ denotes averaging over the random numbers $a$. For fixed walk $\{s_j\}$ and large $N$, the sum $\sum_{j=1}^{N} a_j s_j$ is Gaussian with mean

$$\langle D\rangle = \langle a\rangle \sum_j s_j =: \langle a\rangle M \tag{5}$$

and variance

$$\langle D^2\rangle - \langle D\rangle^2 = N(\langle a^2\rangle - \langle a\rangle^2). \tag{6}$$

The sum over $\{s\}$ is basically an average over all trajectories of our random walk. For large $N$ this average is dominated by trajectories with "magnetization" $M = 0$. Hence the probability of ending the walk at distance $D$ reads

$$p(D) = \frac{1}{\sqrt{2\pi N\langle a^2\rangle}} \exp\left(-\frac{D^2}{2N\langle a^2\rangle}\right). \tag{7}$$

Note that our walker walks on a sublattice of $\mathbb{Z}$ with lattice spacing 2: his movements are confined to the even (odd) numbers for $\sum a_j$ being even (odd). Hence the average number of walks that end at distance $D$ is given by

$$\Omega(D) = 2^N 2p(D) = \frac{2^{N+1}}{\sqrt{2\pi N \langle a^2 \rangle}} \exp\left(-\frac{D^2}{2N\langle a^2\rangle}\right). \quad (8)$$

For the location of the phase transition we can concentrate on perfect partitions, i.e. we assume $D = 0$ and we assume that the $a$'s are uniformly distributed $\kappa N$-bit integers. From

$$\langle a^2 \rangle = \frac{1}{3}\, 2^{2\kappa n}\Big(1 - \mathcal{O}(2^{-\kappa N})\Big) \quad (9)$$

we get

$$\log_2 \Omega(0) = N(\kappa_c - \kappa) \quad (10)$$

with

$$\kappa_c = 1 - \frac{\log_2 N}{2N} - \frac{1}{2N}\log_2\left(\frac{\pi}{6}\right). \quad (11)$$

This is our phase transition: according to (10) we have an exponential number perfect partitions for $\kappa < \kappa_c$, and no perfect partition for $\kappa > \kappa_c$. Our derivation is a bit sloppy, of course, but the result agrees with the rigorous theory of [23].

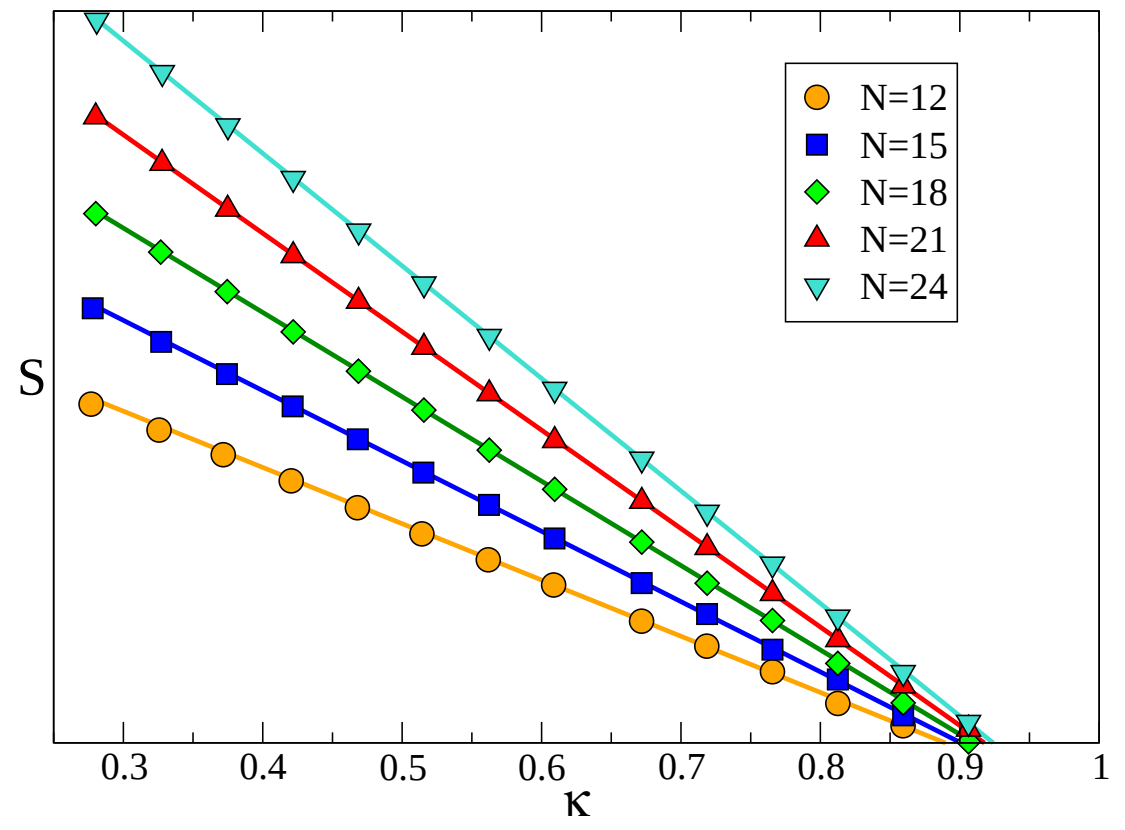

FIG. 3: Entropy $S = \log_2 \Omega(0)$ of perfect partitions vs. $\kappa$. Theory (Eq. (10)) compared to numerical enumerations (symbols).

According to (10) we expect the entropy $S = \log_2 \Omega(0)$ of perfect partitions for fixed but large $N$ to be a linear function of $\kappa$. In fact this can already be observed for rather small problem sizes, see Fig. 3. Linear extrapolation of the simulation data for $\log_2 \Omega(0)$ gives numerical values for the transition points $\kappa_c(N)$. Again the numerical data for small systems agree very well with the predictions of the asymptotic theory (Fig. 4). The strong finite size corrections of order $\log(N)/N$ lead to the curvature of $\kappa_c(N)$ and they are responsible for the incorrect value $\kappa_c = 0.96$ that Gent and Walsh extrapolated from their simulations [21].

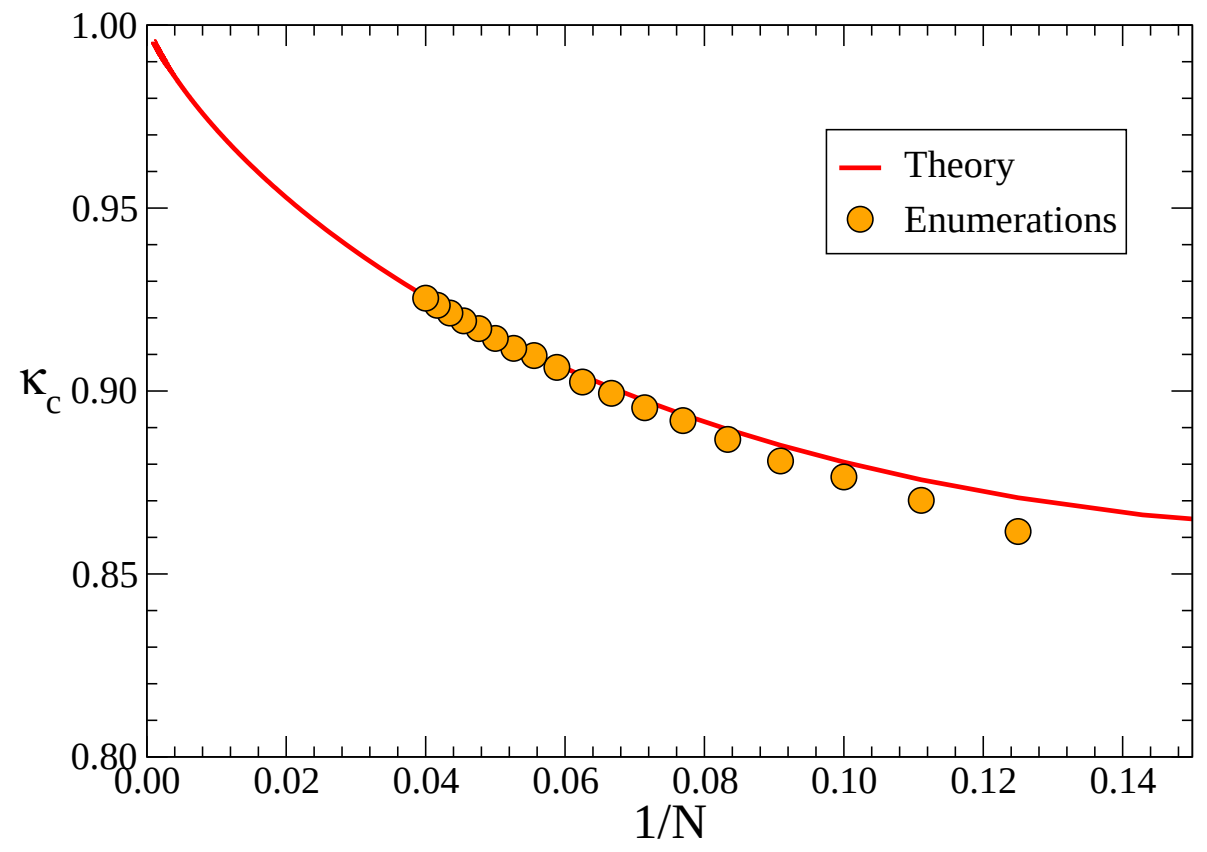

FIG. 4: Numerical data for the transition points $\kappa_c(N)$ have been obtained by linear extrapolation of the data for $\log_2 \Omega(0)$ from Fig. 3. The solid line is Eq. (11).

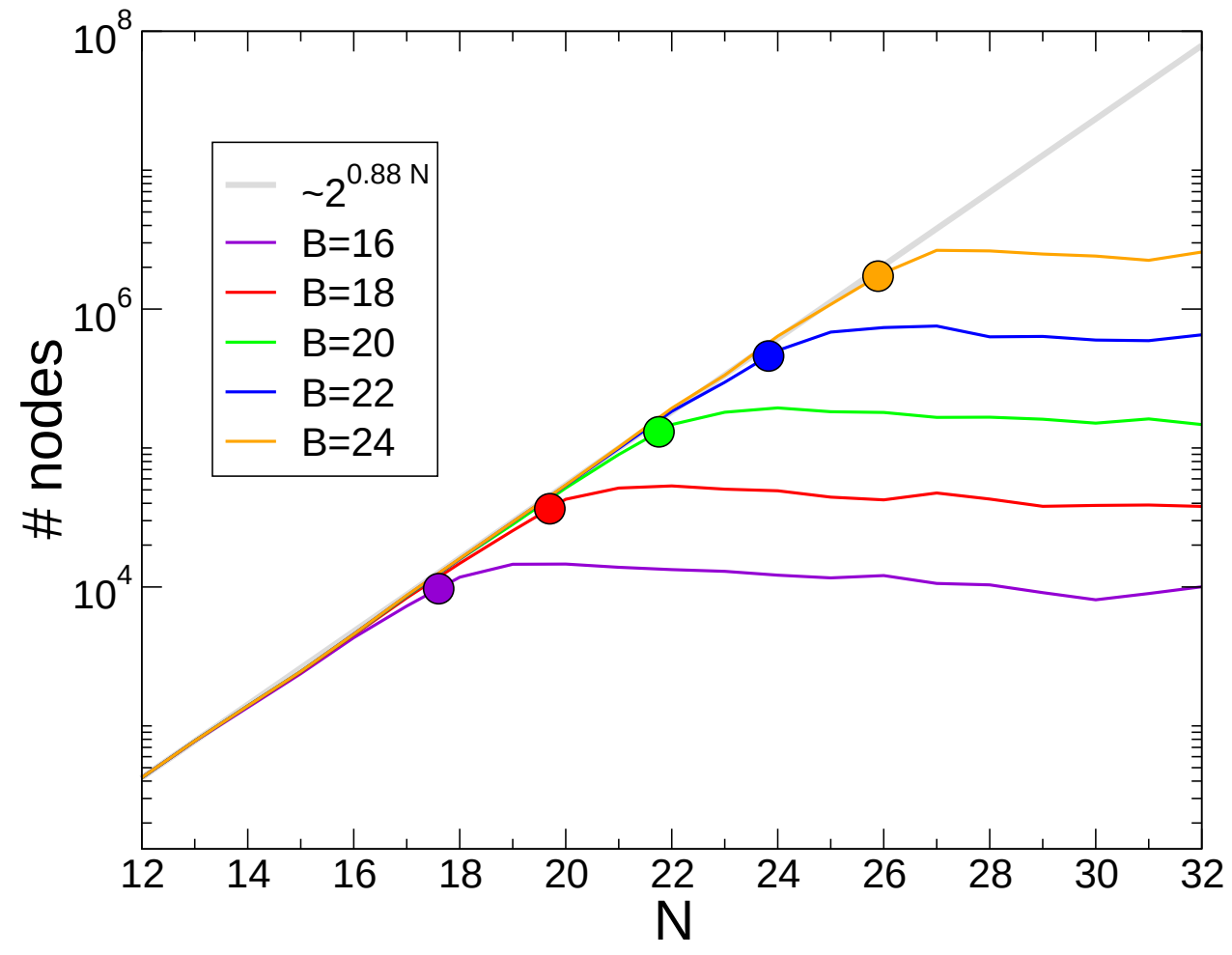

FIG. 5: Partitioning $B$-bit numbers with the complete greedy algorithm: number of search nodes visited vs. $N$. The curves are averages over $10^4$ random samples, the symbols mark the values $N_c$ given by Eq. (12). The fitted curve $2^{0.88N}$ shows that pruning has almost no effect for $N < N_c$.

The phase transition at $\kappa_c$ is a property of the instances. It is by no means clear how this transition affects the *dynamical* behavior of search algorithms. Note that even for $\kappa < \kappa_c$ the fraction of perfect partitions is exponentially small, and finding one of these is non-trivial.

In numerical experiments like the one shown in Fig. 2

the number $B = \kappa N$ of bits is usually fixed and $N$ is varied. Then $\kappa_c$ translates into a critical value $N_c = B/\kappa_c$ or

$$\frac{B}{N_c} = 1 - \frac{\log_2 N_c}{2N_c} - \frac{1}{2N_c}\log_2\left(\frac{\pi}{6}\right). \tag{12}$$

For $B = 20$ this gives $N_c = 21.8$, in good agreement with the location of the hardest instances in Fig. 2. Fig. 5 shows that the average time complexity of the complete greedy algorithm changes its dependence on $N$ precisely at the values $N_c$ given by (12). It is well justified to classify the two regimes $\kappa < \kappa_c$ and $\kappa > \kappa_c$ as "easy" and "hard", resp..

## IV. EASY PHASE

The hallmark of the easy phase is the exponential number of perfect partitions, but the easy phase is not homogeneous: the number of perfect partitions increases with decreasing $\kappa$. This phenomenon *might* yield an interesting structure with regard to algorithms: the performance of an algorithm increases as one moves away from the phase boundary towards smaller values of $\kappa$. In fact, Fig. 2 indicates that complete differencing finds a perfect partition with its very first descent in the search tree if $\kappa$ is small enough. Maybe the easy phase disintegrates into two phases, one in which complete differencing has to backtrack, and another one in which the first try hits a perfect solution?

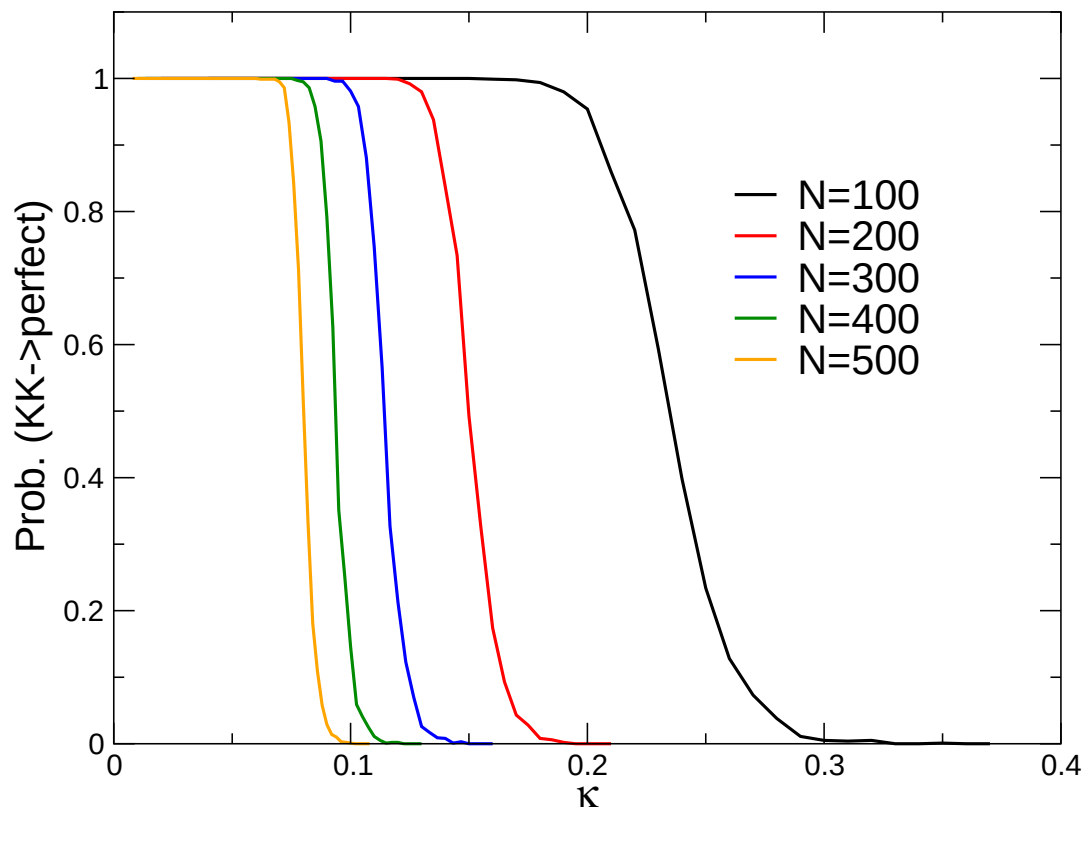

FIG. 6: Probability of the event "Karmarkar-Karp differencing heuristics yields perfect partition".

To check this hypothesis we investigate the Karmarkar-Karp (KK) heuristic solution of the NPP. Remember that this solution is the first one generated by the complete differencing algorithm. Let $D_{kk}$ be the discrepancy of the KK-solution. Our hypothesis then is: There is a value $0 \leq \kappa_{kk} \leq \kappa_c$ such that

$$\lim_{N\to\infty} \text{Prob.}(D_{kk} \leq 1) = \begin{cases} 1 & \kappa < \kappa_{kk} \\ 0 & \kappa > \kappa_{kk} \end{cases} . \tag{13}$$

Fig. 6 shows the result of a simulation of the KK-algorithm. In fact there is a sharp transition at a value $\kappa_{kk}$, but this value depends on $N$ and seems to go to 0 as $N \to \infty$.

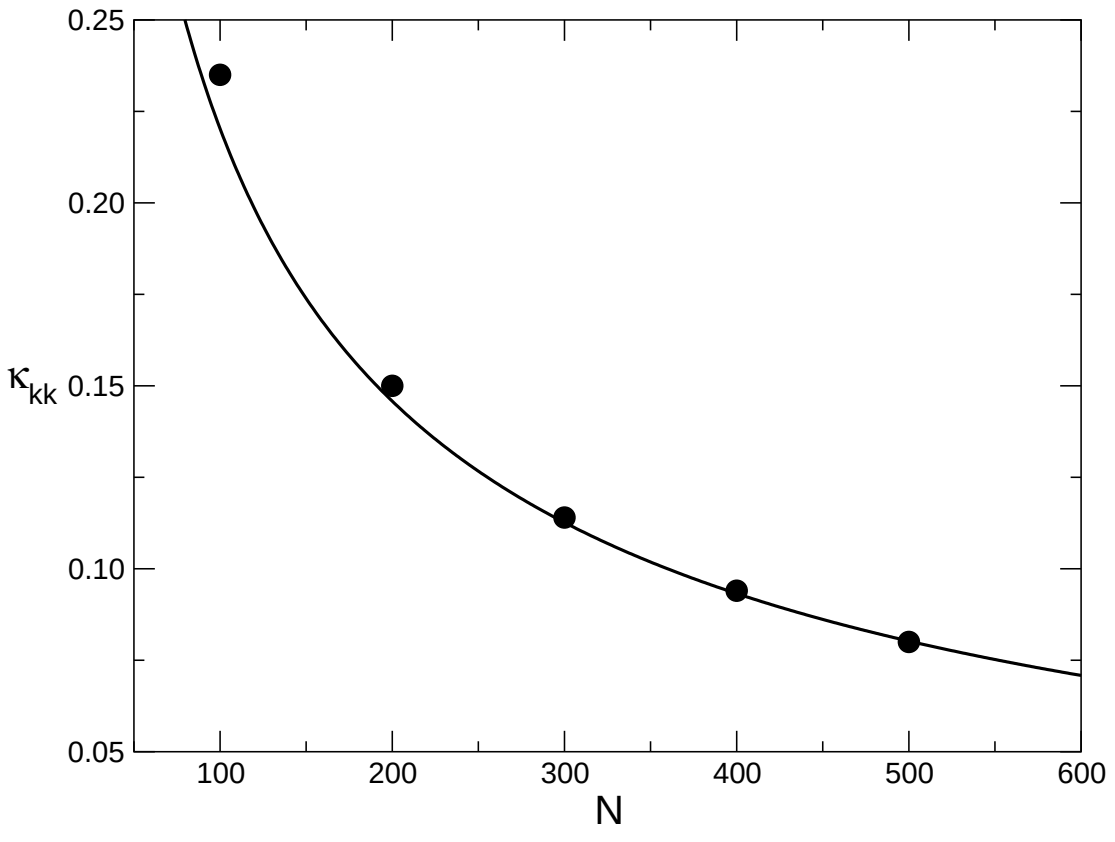

FIG. 7: Threshold value $\kappa_{kk}$ below which Karmarkar-Karp differencing yields perfect solutions. Eq. 16 (solid line) compared to numerical simulations (symbols).

A simple consideration points out how this happens: We know from the work of Yakir [13] that the KK-algorithm generates partitions with discrepancy

$$D_{kk} = N^{-\alpha \log N} \tag{14}$$

for some constant $\alpha > 0$. For a perfect partition, all $\kappa N$ bits of the discrepancy must be zero, or

$$N^{-\alpha \log N} \leq 2^{-\kappa N}. \tag{15}$$

This inequality is fulfilled as long as $\kappa \leq \kappa_{kk}(N)$ with

$$\kappa_{kk}(N) = \alpha \frac{\log^2 N}{N \log 2} \tag{16}$$

Fig. 7 shows $\kappa_{kk}(N)$ in comparison with the results from simulations, where we measured $\kappa_{kk}$ as the value where the probability of generating a perfect partition is 1/2. For $\alpha$ we took the value 0.72 reported by Yakir for the average discrepancy of the KK-solution. Yakir's proof of eq. 14 was based on the continuous case $\kappa = \infty$, but it is probably not too hard to extend it to prove (16).

Note that a similar consideration indicates that even the greedy heuristic eventually yields perfect partitions for very small values of $\kappa$. For the parameter $B = 20$ used in Fig. 2 we expect the greedy heuristics to generate perfect partitions for $N > 839000$.

## V. HARD PHASE

Fig. 2 shows that in the easy phase complete differencing outperforms complete greedy, and in view of the exponentially small fraction of perfect partitions both algorithms outperform blind search through all partitions. Fig. 2 also indicates that complete greedy and complete differencing are coequal in the hard phase. In fact, in the hard phase both are coequal to blind random search, as we will see in this section.

A first hint on the hardness of the NPP in its hard phase was provided by the *random cost approximation* to the NPP [28]. Here the original problem is replaced by the problem to locate the minimum number in an unsorted list of $2^{N-1}$ *independent* random, positive numbers $E$ drawn from the density

$$p(E) = \frac{2}{\sqrt{2\pi N \langle a^2 \rangle}} \exp\left(-\frac{E^2}{2N\langle a^2\rangle}\right) \qquad (E \geq 0). \tag{17}$$

This is the probability density of discrepancy in the NPP, confer (7), but of course the discrepancies in the NPP are not independent random variables. The approximation of independence on other hand allows the calculation of the statistics of the optimal and near optimal solutions. Consider the continuous case, i.e. $\kappa \to \infty$. Then the numbers $E$ are real, positive numbers drawn from (17). Let $E_k$ denote the $k$-th lowest of these numbers. The probability density $\rho_1$ of the minimum $E_1$ can easily be calculated:

$$\rho_1(E_1) = 2^{N-1} \cdot p(E_1) \cdot \left(1 - \int_0^{E_1} p(E')dE'\right)^{2^{N-1}-1} \tag{18}$$

$E_1$ must be small to get a finite right-hand side in the large $N$ limit. Hence we may write

$$\begin{aligned}\rho_1(E_1) &\approx 2^{N-1} \cdot p(0) \cdot \left(1 - E_1 p(0)\right)^{2^{N-1}-1} \\ &\approx 2^{N-1} \cdot p(0) \cdot e^{-2^{N-1}p(0)E_1}.\end{aligned}$$

This means that the scaled minimum,

$$\varepsilon_1 = 2^{N-1} \cdot p(0) \cdot E_1 \tag{19}$$

is an exponential random variable,

$$\rho_1(\varepsilon) = e^{-\varepsilon} \qquad (\varepsilon > 0) \tag{20}$$

Along similar lines [29] one can show that the density $\rho_k$ of the $k$-th lowest scaled number is

$$\rho_k(\varepsilon) = \frac{\varepsilon^{k-1}}{(k-1)!} \cdot e^{-\varepsilon} \qquad k = 2, 3, \ldots. \tag{21}$$

Figs. 8 and 9 compare Eqs. (20) and (21) with the probability density of the scaled optimal and near-optimal discrepancies in the NPP. The agreement is amazing, even for small values of $N$. In fact (20) and (21) have been established as the asymptotic probability measure for the optmimum discrepancies—rigorously and without the assumption of independence [23].

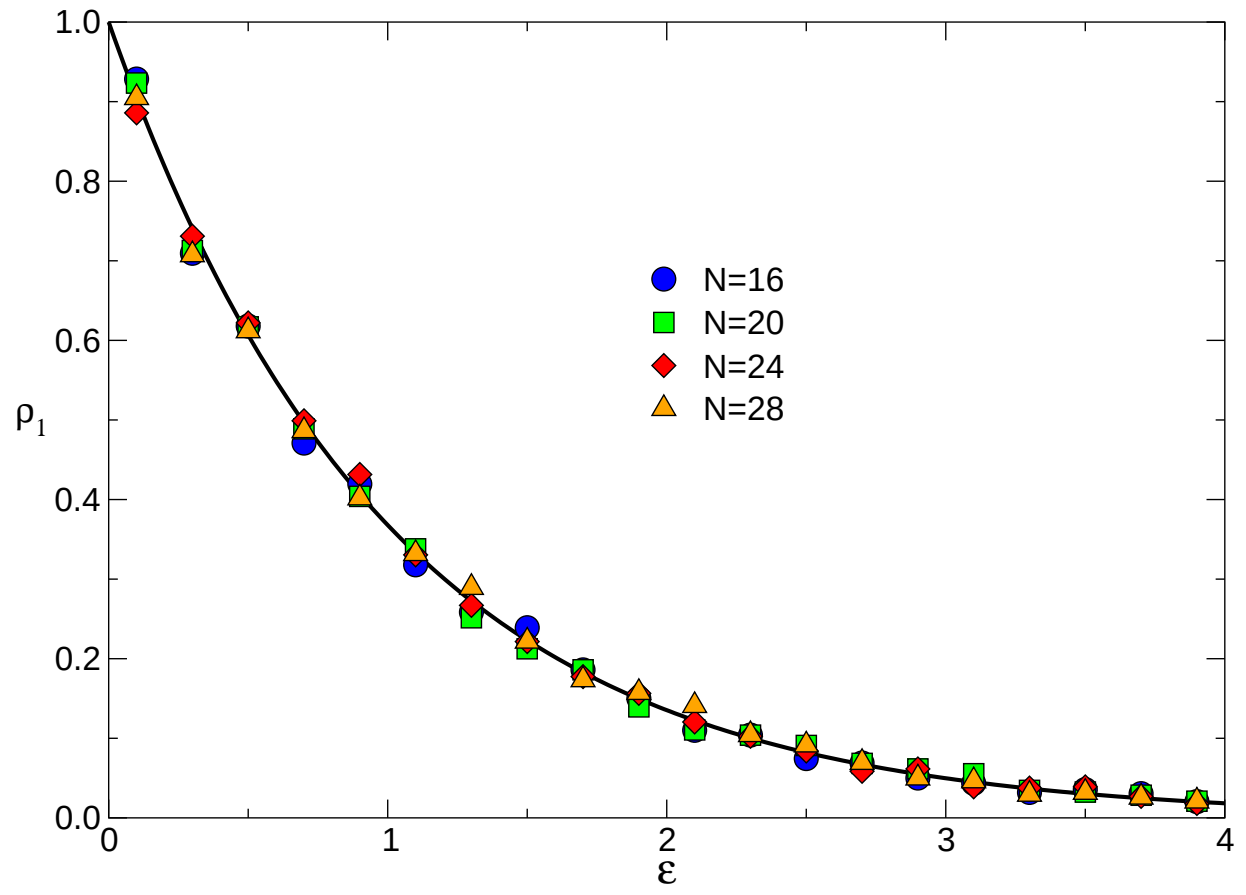

FIG. 8: Probability density of the scaled optimum discrepancy in the hard phase. Symbols: Numerical simulations. Solid line: prediction by the random cost approximation.

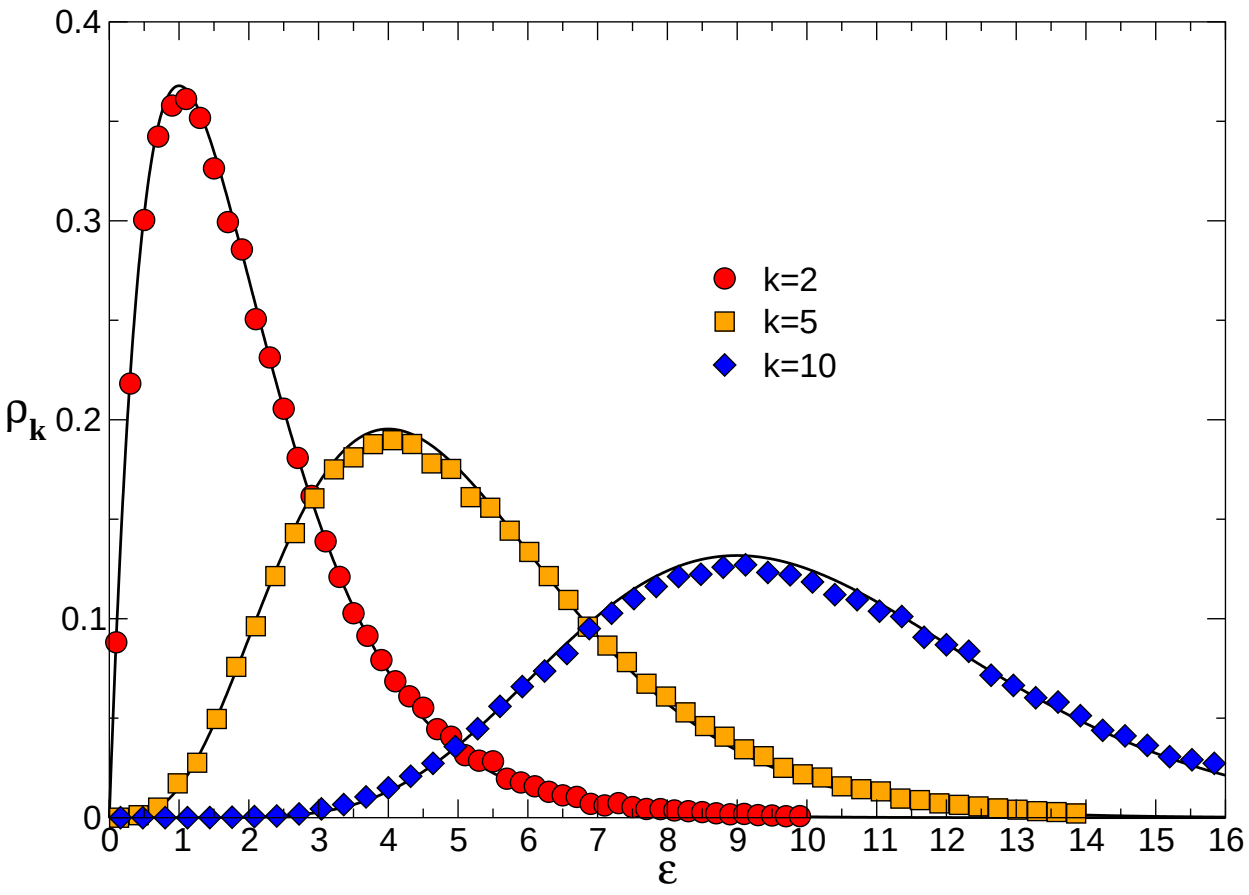

FIG. 9: Probability densities of the $k$-th best partition in the hard phase. Symbols: Numerical simulations for $N = 24$. Solid lines: predictions by the random cost approximation.

The fact that the random cost approximation gives the accurate statistics of the optimum discrepancies is no accident, of course. There is a certain degree of statistical independence among the costs in the NPP. This can be seen from the joint probability

$$p(E, E') = 2^{-2N} \sum_{\{s_j, s'_j\}} \left\langle \delta(E - |\sum_j a_j s_j|)\, \delta(E' - |\sum_j a_j s'_j|) \right\rangle \tag{22}$$

of finding discrepancies $E$ and $E'$ in one instance of an NPP. In [30] it is shown that this probability factorizes, i.e. $p(E, E') = p(E)p(E')$ for discrepancies $E$ and $E'$ that are smaller than $\mathcal{O}(N)$. To depict the uncorrelatedness of the small discrepancies consider a partition in the continuous problem with very low discrepancy $E$ of order $\mathcal{O}(\sqrt{N}\,2^{-N})$. Any single local move $s_j \mapsto s'_j = -s_j$ increases $E$ by something $\mathcal{O}(N^{-1})$, and it takes a lot of moves to compensate this to get another discrepancy $E' = \mathcal{O}(\sqrt{N}\,2^{-N})$. The corresponding partitions $s$ and $s'$ typically have an overlap 0, and this leads to the factorization of $p(E, E')$.

The random cost problem is an algorithmic nightmare. No smart heuristic is better than stupid random or sequential search, and this is the reason why there are no good heuristics in the hard phase of the NPP, and why complete algorithms cannot really take advantage of the pruning rules. But there are differences in the quality of heuristic solutions, remember the greedy result ($\mathcal{O}(N^{-1})$) and the KK-heuristic ($\mathcal{O}(N^{-\alpha \log N})$). How can these differences arise if the NPP is essentially a random cost problem? The answer is that both algorithms exploit the correlations among the *large* discrepancies to stay away from the bad partitions, and the differencing method is much more efficient at this. The correlations between large discrepancy configurations are also responsible for the fact that the complete barrier tree of the NPP looks different from the complete barrier tree of the pure random cost problem [31].

Complete algorithms differ only in the sequence in which they explore the partitions. In the sequence generated by complete differencing the true optimum might appear earlier than in the sequence generated by complete greedy, but if the random cost picture is correct, the location of the optimum is random in any prescribed sequence. This has been checked for example for another smart algorithm proposed by Korf [27]. Korf suggested to reorder the leafs of the search tree of the complete differencing method according to their number of "right turns" (violations of the differencing heuristics) in their pathes, starting with those leafs that deviate less from the KK-heuristic. In our example from Fig. 1 the leafs would be visited in order $(2, 4, 4, 6, 0, 6, 8, 14, 8, 10, 12, 16, 18, 20, 22, 30)$, and in fact the perfect solution appears earlier than in the sequence shown in Fig. 1. Numerical simulation however revealed that in the hard phase the position of the optimum in the sequence generated by this algorithm is completely random, as predicted by the random cost problem [30].

Apparently there is no way to overcome the random cost nature of the NPP in the hard phase, or as Brian Hayes put it, "When the NPP is hard, it's very hard."

## VI. CONCLUSIONS

We have seen that random NPP has a phase transition in average complexity, and that this phase transition goes hand in hand with a transition in probability of perfect solutions. The control parameter $\kappa$ of both transitions is the ratio of the number of bits in the $a_j$'s and the number $N$ of variables, and $\kappa_c = 1 - \log_2(N)/2N + \mathcal{O}(N^{-1})$ is the critical value that separates the hard ($\kappa > \kappa_c$) from the easy ($\kappa < \kappa_c$) phase. Much more can be said about the phase transition, for example about the width of the transition window and the probability of perfect solutions inside that window. Another proven fact is the uniqueness of the solution in the hard phase. All this (and much more) can be found in the paper of Borgs, Chayes and Pittel [23]. Their work answers most of the open questions on random NPP that are not related to algorithms. The major open problem is to put the random cost approximation on rigorous grounds and to clarify its relevance for algorithms. From a practical point of view it would be very nice to have a polynomial time algorithm that yields better results than the differencing method. After all, there is much room between $\mathcal{O}(N^{-\alpha \log N})$ and $\mathcal{O}(\sqrt{N}\,2^{-N})$.

The NPP as shown here can be generalized and modified in various directions. An obvious generalization is to partition the numbers into $q \leq 2$ subsets. This is called the MULTIPROCESSOR SCHEDULING PROBLEM, and in physics parlance this corresponds to a Potts spin glas or to a walk with random stepszies in $q-1$ dimensions. The latter approach has been used to analyze an "easy-hard" phase transition in MULTIPROCESSOR SCHEDULING [7].

Another variant is the constrained NPP where the cardinality of the subsets is fixed. This is necessary for some appications like choosing up sides in a ball game [4], where both teams need to have the same number of players. The cardinality difference of the subsets is a control parameter that triggers another phase transition in computational complexity, giving rise to a 2-dimensional phase diagram [32].

### Acknowledgments

Discussions with Heiko Bauke are gratefully acknowledged. Part of the numerical simulations have been done on our Beowulf cluster TINA[33].